\documentclass[3p,times,twocolumn]{elsarticle}

\usepackage{ecrc}

\volume{00}

\firstpage{1}

\journalname{Nuclear Physics B Proceedings Supplement}

\runauth{B. Bilin for the CMS Collaboration}

\jid{nuphbp}

\jnltitlelogo{Nuclear Physics B Proceedings Supplement}

\usepackage{amssymb}

\usepackage[figuresright]{rotating}

\begin{document}

\begin{frontmatter}



\dochead{}

\title{Differential Z + jet cross section measurements at 8 TeV}


\author{Bugra Bilin on behalf of the CMS Collaboration}

\address{Middle East Technical University, CERN PH-UCM Bat 42 2-029 C28810, CH 1211 Geneva 23, Switzerland}

\begin{abstract}
The measurement of differential cross section of a Z boson produced in association with jets is presented. The cross section is presented with respect to various jet kinematic variables where the Z bosons are reconstructed from opposite sign lepton pairs. The analysis is based on data of proton proton collisions with the centre of mass energy of 8 TeV collected in 2012 by the CMS experiment at LHC corresponding to 19.8 fb$^{-1}$ of integrated luminosity. Obtained results are compared with different  generators and are shown to be consistent with the Standard Model predictions.
\end{abstract}


\end{frontmatter}


\section{Introduction}
\label{sec:intro}

The large center-of-mass energy of pp collisions at the LHC allows the production of events with high jet transverse momentum, $p_T$, and high number of jets, $N_{jet}$, with a Z boson. Z bosons decaying to opposite sign lepton pairs (e, $\mu$)  provide an almost background free signal. Measurements of these processes provide stringent tests of pQCD predictions, and they are backgrounds to many Standard Model (SM) measurements such as single top, tt, vector boson fusion, WW scattering, Higgs boson production, as well as Beyond the Standard Model (BSM) searches such as supersymmetry (SUSY). 

CMS collaboration \cite{CMS} reported the first cross section measurements of Z boson production in association with jets  (Z + jets)  \cite{SMP-13-007,SMP-14-009} at a center-of-mass energy of $\sqrt{s}$ = 8 TeV using full 2012 data of 19.6 fb$^{-1}$ of integrated luminosity. In sections \ref{sec:double} and \ref{sec:single} the double and single differential Z + jet cross section measurements are given respectively. The cross sections are presented after deconvoluting the detector effects by an unfolding procedure. The measurements presented cover several kinematic regions in jet rapidity, and hence provide stringent tests of perturbative QCD predictions merging parton shower with matrix element calculations at leading (LO+PS) and next-to-leading (NLO+PS) order. They also provide tests of the Parton Distribution Functions (PDFs) mainly at these large rapidities.

\section{Double Differential Z+jet Cross Section Measurements}
\label{sec:double}

The double differential cross section, $d^2\sigma/dp_{T}^jdy^j$, is measured with respect to the transverse momentum $(p_{T})$ and the rapidity ($y$) of the highest $p_{T}$ jet in the di-muon channel. The muons are required to have $p_{T}$ greater than 20 GeV and pseudo-rapidity ($\eta$) less than 2.4. Particle-Flow \cite{PF} jets are selected in the calorimeter acceptance of $|y|<4.7$ using the anti-$k_{T}$ jet clustering algorithm with a cone size of $\Delta R =0.5$. The jets are required to  have $p_{T} > 30$ GeV for jets in the region $|y|$ $<$ $2.5$, and $p_T > 50$ GeV for $|y|>2.5$. The jets with $\Delta R(j,\mu)<0.5$ are not considered in the analysis.

\begin{figure}
\includegraphics[width=8.6cm,clip]{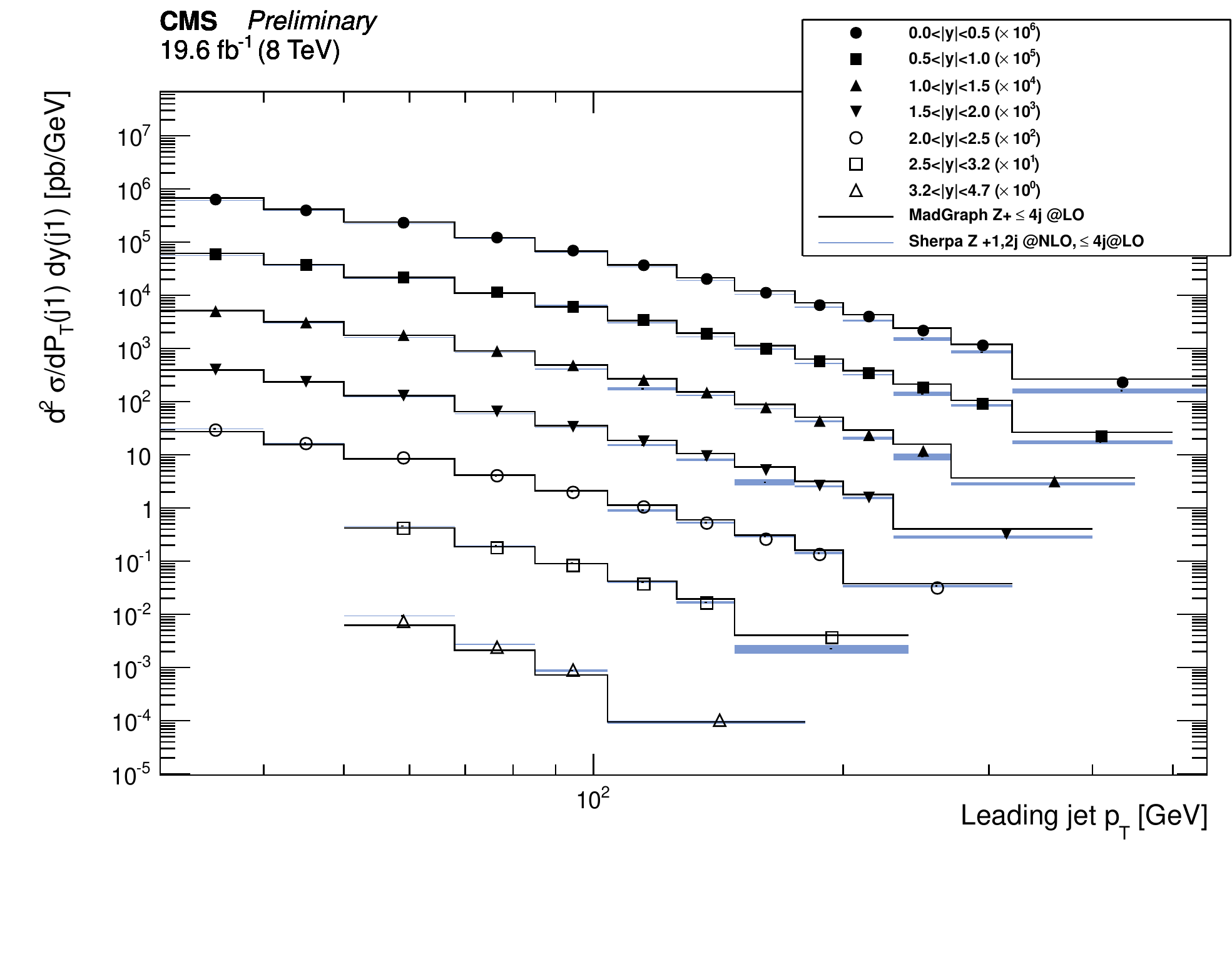}
\caption{Double differential cross section versus leading jet transverse momentum for various rapidity bins in the di-muon channel. Data points are shown with statistical error bars. The black lines are  \textsc{MadGraph} predictions normalised to the inclusive NNLO cross section. The \textsc{sherpa2} predictions is shown as blue band, whose thickness indicates the statistical uncertainty.}
\label{fig:twod}       
\end{figure}

\begin{figure}
        \includegraphics[width=7.6cm]{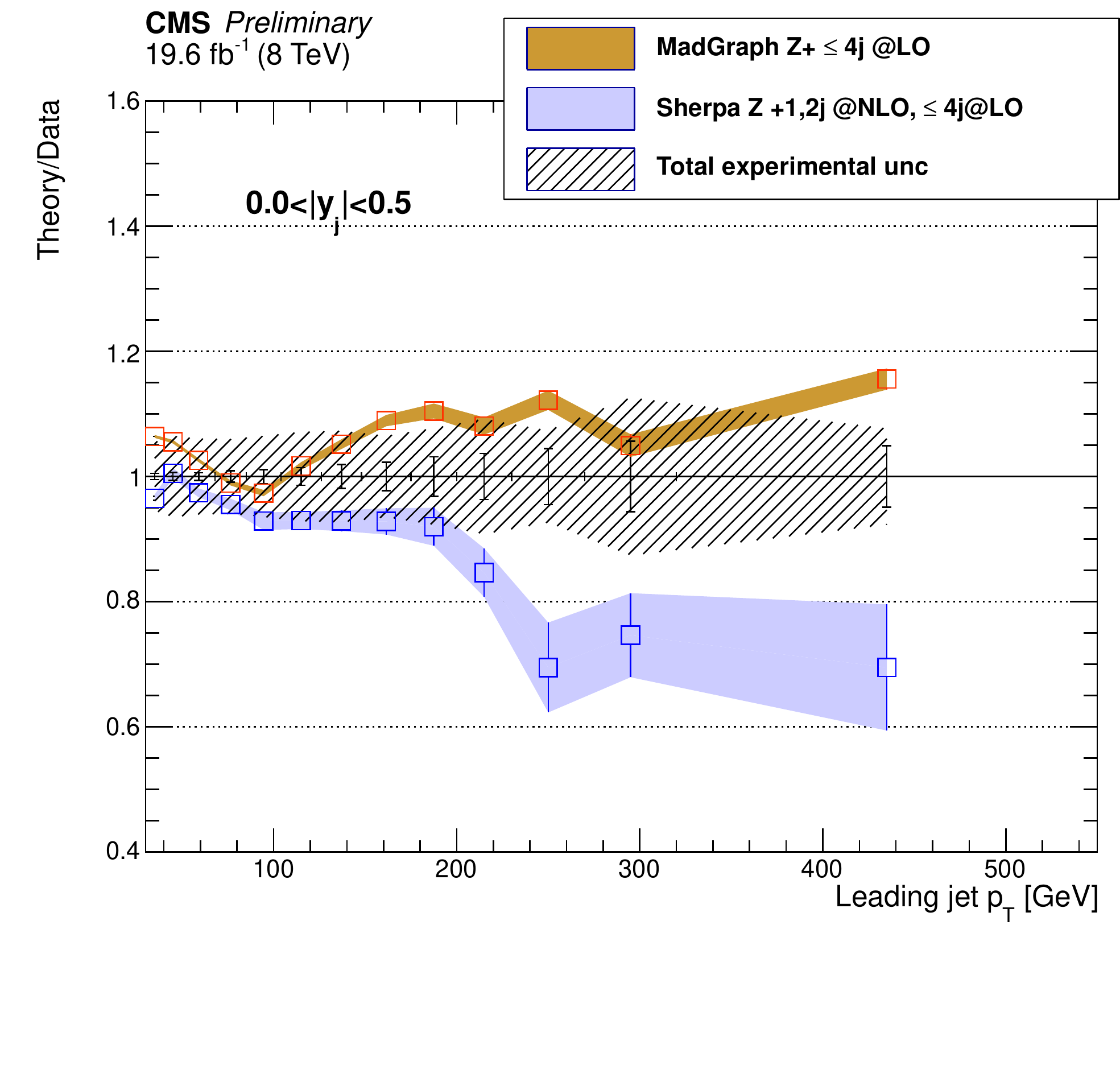}
         \caption{Ratio of theoretical predictions to the data for double differential cross section versus leading jet transverse momentum for the first rapidity bin in the di-muon channel. The statistical uncertainty of the measurement is presented by error bars located around the theory/data = 1 line. The quadratic sum of experimental uncertainties and the statistical uncertainty of the measurement is shown as a shaded band. Error bars on the ratios represent the statistical uncertainties of the MC predictions.}
      \label{fig:twodratio}
\end{figure}

In figure \ref{fig:twod} and \ref{fig:twodratio}, the double differential cross section results versus leading jet $p_T$ and the ratio of theory predictions to measurements are given respectively. 

 \section{Single Differential Z+jet Cross Section Measurements}
\label{sec:single}
Single differential cross section of Z + jet process is measured in the di-electron and di-muon final states. The measurements are carried out with respect to the exclusive and inclusive jet multiplicity, jet $p_T$, jet $|\eta|$ and jet transverse momentum scalar sum ($H_T$) .
The leptons are required to have $p_{T}$ greater than 20 GeV and $|\eta|$ less than 2.4. A threshold of $30$ GeV on jets $p_T$ is applied to reduce the pileup contamination as well as the large jet energy correction uncertainty. The jets overlapping with the leptons with $\Delta R(j,l)<0.5$ are not considered in the analysis.

\begin{figure}
\includegraphics[width=8.7cm,clip]{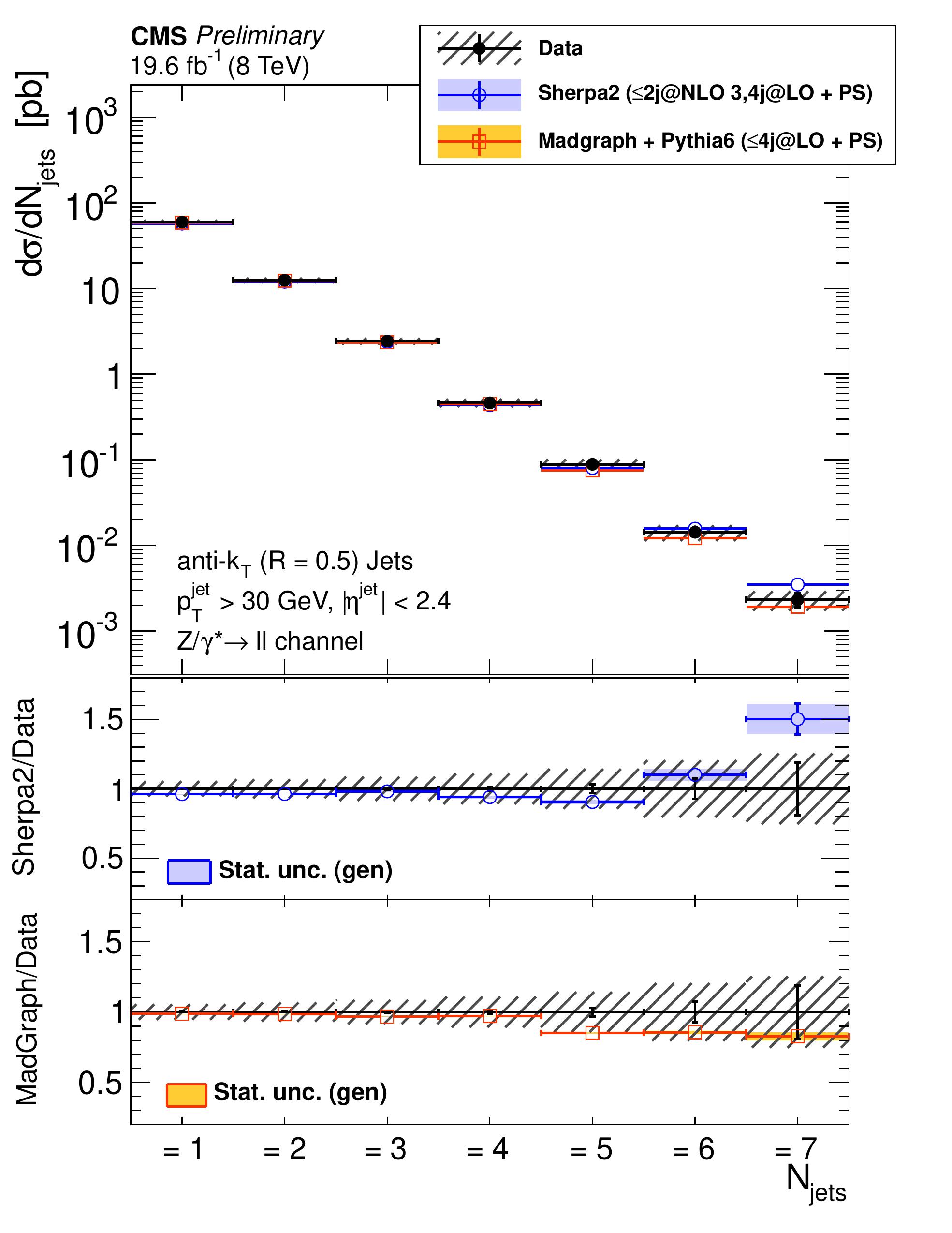}
\caption{Differential cross section measured as a function of the exclusive jet multiplicity distribution compared to the  \textsc{sherpa2} and  \textsc{MadGraph} Monte Carlo predictions. The lower panels show the ratio of theory prediction to data. Error bars around the experimental points show the statistical uncertainty, while the crosshatched bands indicate the statistical plus systematic uncertainties added in quadrature. The colored filled band around theory represents the statistical uncertainty of the generated sample.}
\label{fig:mult}       
\end{figure}

\begin{figure}
\includegraphics[width=8.7cm,clip]{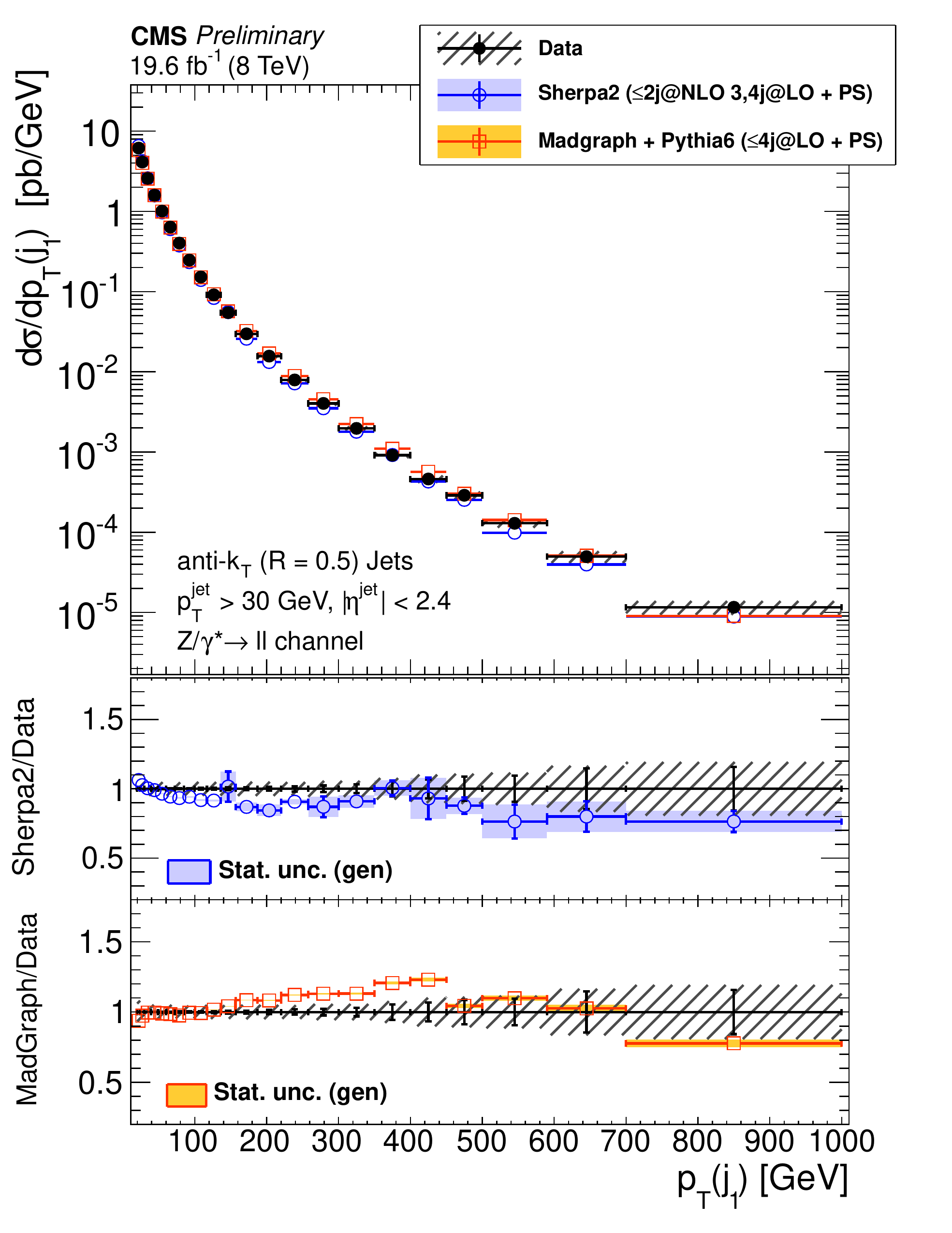}
\caption{Differential cross section measured as a function of of the 1st jet $p_{T}$ compared distribution compared to the  \textsc{Sherpa2} and  \textsc{MadGraph} Monte Carlo predictions. The lower panels show the ratio of theory prediction to data. Error bars around the experimental points show the statistical uncertainty, while the crosshatched bands indicate the statistical plus systematic uncertainties added in quadrature. The colored filled band around theory represents the statistical uncertainty of the generated sample.}
\label{fig:fivejpt}       
\end{figure}

Cross section results as a function of exclusive jet multiplicity and the 1st jet $p_T$ distributions are shown in figures  \ref{fig:mult} and \ref{fig:fivejpt} respectively.

\section{Conclusion}
\label{sec:concl}

 Single differential cross section measurements  have been made as a function of exclusive and inclusive jet multiplicities, as a function of the $p_T$ and $\eta$ of the $n^{\rm{th}}$ jet for $n = 1\dots5$, and as a function of the scalar sum of the jet transverse momenta for $N_{\rm{jets}} \geq n$, for $n = 1\dots5$. Also, the double differential cross section measurement with respect to the rapidity and transverse momentum of the highest $p_T$ jet is presented in the di-muon final state, which is the first study of the double differential cross section in the Z + jet final state, and the first CMS measurement including forward jets up to $|y|<4.7$ in this Z +  jet event topology.

The measurements have been compared with two different calculations with different fixed order accuracy. Comparison is done with respect to tree level predictions from \textsc{MadGraph} \cite{MG}, and \textsc{sherpa2} \cite{SH}. \textsc{MadGraph} is a tree level matrix element calculator which generates Z bosons associated with up to four partons. \textsc{pythia6}  \cite{PHY} is used to add the remaining QCD radiation via parton showering algorithm. \textsc{sherpa2} is a multi-leg NLO generator. Events with up to two partons along with the Z boson are generated at NLO and merged with LO matrix element calculations up to configurations with a Z and four partons.

An excess of the cross-section contribution in the region $p_{T}^{j_1}$ $\approx$ 150 $-$ 450 GeV with respect
to the rest of the phase space in the \textsc{MadGraph} + \textsc{pythia6} calculations is observed in both measurements, whereas  \textsc{sherpa2}
calculations predict a slightly harder spectrum than the measurement. An overall agreement is seen between \textsc{sherpa2} predictions and the data, except some discrepancies in different y and $p_{T}$ regions in the double differential cross section measurements. On the overall scale, the measurements are in agreement with the theoretical predictions within uncertainties. 







\end{document}